\documentclass[12pt,a4paper]{article}
\usepackage[english]{babel}
\usepackage[utf8]{inputenc}
\usepackage{amsmath}
\usepackage{amsfonts}
\usepackage{amssymb}
\usepackage{graphicx,color}
\usepackage{slashed}
\usepackage{physics}
\usepackage{tikz}
\usepackage{pgfplots}
\usepgfplotslibrary{polar}
\pgfplotsset{compat=1.17}
\begin{document}
\title{Spatial dependence of quantum friction amplitudes in a scalar
model} 
\author{Aitor Fern\'andez and C.~D.~Fosco\\
{\normalsize\it Centro At\'omico Bariloche and Instituto Balseiro}\\
{\normalsize\it Comisi\'on Nacional de Energ\'{\i}a At\'omica}\\
{\normalsize\it R8402AGP S.\ C.\ de Bariloche, Argentina.} }
\maketitle
\begin{abstract} 
We study the spatial dependence of the  quantum friction effect for an atom
moving at a constant velocity, in a parallel direction to a material plane.
In particular, we determine the probability per unit time and unit area,
for exciting degrees of freedom on the plane, as a function of their 
position, for a given trajectory of the atom.
We also show that the result of integrating out the probability density
agrees with previous results for the same system.
\end{abstract}
\section{Introduction}\label{sec:intro} 
The intrinsically quantum nature of the elementary constituents of matter
and their interactions can sometimes manifest itself macroscopically, in
a rather straightforward way. Indeed, among the most distinctive features
of quantum systems are their vacuum fluctuations, which produce
observable effects when subjected to non-trivial boundary
conditions.  This is the case in the celebrated Casimir
effect~\cite{libros}, where material media imposes boundary conditions on
the electromagnetic (EM)  field fluctuations. 

A different kind of phenomenon, where quantum fluctuations are also
responsible of observable effects, is the so-called ``non-contact
friction'' or ``Casimir friction'', whereby a frictional force appears on
lossy media in non-accelerated relative motion. It is a somewhat
complementary situation to the Casimir effect case, since the zero point
fluctuations of the EM field are not directly relevant; rather, its role is
to mediate the interaction between the microscopic degrees of freedom on
the two media.  The frictional effect does not happen for perfects
mirrors~\cite{Pendry97} but if may appear in non-dispersive
media~\cite{Maghrebi:2013jpa} when their relative speed overcomes the
threshold posed by the speed of light in the media.  
The dissipative force may also appear on a single atom moving with constant
velocity, parallel to a plate~\cite{others}.
There are also thresholds related to the speed of the modes on the material
media; for instance, in a recent paper~\cite{Fosco:2021aih}, for an atom
in the proximity of a graphene plate the atom must move faster than $v_F$, the Fermi
speed of the electrons in graphene, for dissipation to occur.

In this paper, we use a quantum field
theory model to study quantum friction between an atom, moving at a constant 
parallel speed with respect
to a planar medium, in an approach which allows us to study the spatial
distribution of the media excitations which play a role for the existence
of the frictional force. 

The model we use is essentially the same we  had used in~\cite{Farias:2019lls}, which is based on
\cite{Galley:2012qz},  namely, a  vacuum scalar field
linearly coupled to a set of uncoupled quantum harmonic oscillators which
are the microscopic ``matter'' degrees of freedom on the mirror.
Note that when considering the quantum friction effect between two planes,
the spatial details we want to study are lost because of the very geometry
of the system.

Here, we use a perturbative quantum field theory approach to calculate
transition amplitudes, and those amplitudes account for processes whereby
modes on the medium are excited, what allows us to study the spatial
distribution of the phenomenon. Besides, by integrating out the transition
probabilities, we also provide an indirect verification of the result
obtained in~\cite{Farias:2019lls}, where the total probability of vacuum
decay was otained from the imaginary part of the in-out effective action to
the frictional force on the plates. 
A similar approach has been used in~\cite{Fosco2011} and~\cite{Fosco:2007nz}
while in~\cite{BelenFarias:2014ehx} a CTP in-in formulation~\cite{CTP} has
been applied to evaluate the frictional force between two plates in
relative motion at a constant speed.

\section{The system}\label{sec:thesystem}
The system we deal with here is, regarding both its dynamical variables and 
the interactions between them, essentially the same as the one considered
in~\cite{Farias:2019lls}. It contains a scalar variable $q$, associated with
the ``electron'': a scalar degree of freedom sitting on a moving atom, while 
the atom's center of mass trajectory, ${\mathbf r}(t)$, is externally driven. 
The  variable $q$ is coupled to a vacuum real scalar field
$\varphi$, which also interacts with a medium, represented by microscopic
independent scalar degrees of freedom $Q$, uniformly distributed on a plane.

Regarding conventions, in this paper we shall use natural units, so that 
$c = 1$ and $\hbar = 1$; space-time coordinates are denoted 
by $x = (x^\mu)_{\mu=0}^3$, $x^0 =  t$, and we use the Minkowski metric
$(g_{\mu\nu}) \equiv {\rm diag}\{ 1,-1,-1,-1 \}$. Our choice of coordinates is
such that the spacetime occupied by the medium is $x^3 = 0$. 
Correspondingly, space-time coordinates relevant to the degrees of freedom on 
the plane, shall be denoted by \mbox{$x_\shortparallel = (x^\alpha)_{\alpha=0}^2 = (t,{\mathbf
x}_\shortparallel)$}. Here, ${\mathbf x}_\shortparallel \equiv (x^1,x^2)$
are two Cartesian coordinates on the spatial plane.  

The real-time action ${\mathcal S}$ for the whole system will thus be
conveniently defined as follows:
\begin{equation}
	{\mathcal S}(q, Q, \varphi; {\mathbf r}(t)) \;=\; 
	{\mathcal S}^{(0)}(q, Q, \varphi) 
	\,+\, {\mathcal S}'(q, Q, \varphi; {\mathbf r}(t)) \;,
\end{equation}
where ${\mathcal S}^{(0)}$ determines  the free evolution, while 
${\mathcal S}'$ does so for the interactions. The free part will consist of
three terms: ${\mathcal S}^{(0)}_e$ for the electron,  ${\mathcal
S}^{(0)}_m$ for the medium, and ${\mathcal S}^{(0)}_v$ for the vacuum
field:
\begin{equation}
{\mathcal S}^{(0)}(q, Q, \varphi) \,=\, 
{\mathcal S}^{(0)}_e(q) \,+\, 
{\mathcal S}^{(0)}_m(Q) \,+\, 
{\mathcal S}^{(0)}_v(\varphi) \;,
\end{equation}
where:
\begin{equation}
{\mathcal S}^{(0)}_e(q) \,=\, \int dt \,\frac{m}{2}  
\big( \dot{q}^2(t)  \,-\, \Omega_e^2
q^2(t) \big) \;,
\end{equation}
\begin{equation}
{\mathcal S}^{(0)}_v(\varphi) \,=\, \int d^4x \, \frac{1}{2} 
\partial_\mu \varphi(x) \partial^\mu \varphi(x) \;,
\end{equation}
and
\begin{equation}
{\mathcal S}^{(0)}_m(Q) \,=\, \int d^3x_\shortparallel \, \frac{1}{2} 
\left[  \partial_t Q(x_\shortparallel)  \partial_t Q(x_\shortparallel)  
\,-\, \Omega_m^2  Q^2(x_\shortparallel)
\right] \;,
\end{equation}
where $m$ is the mass of the electron's degree of freedom. It is assumed that
its free dynamics is the one of a harmonic oscillator with the frequency $\Omega_e$ determining
its energy levels. As we shall see, at the lowest non-trivial order, the physics of the
quantum friction process is determined by the ground state and the first excited state.
Thus, results should be in this respect quite universal regarding the potential for $q(t)$, 
except for a redefinition of the parameters; for example, the energy gap.

On the other hand, note that the medium may be thought of as a continuous
distribution of decoupled oscillators with frequency $\Omega_m$. It
corresponds to taking the $u \to 0$ limit in a more general model, namely,
one whose elementary excitations have speed $u$:
\begin{equation}
	{\mathcal S}^{(0)}_m(Q) \,=\, \lim_{u \to 0}
	\int d^3x_\shortparallel \, \frac{1}{2} 
	\left[  \big(\partial_t Q(x_\shortparallel) \big)^2 \,-\, u^2 \,
	\big| {\mathbf \nabla}_\shortparallel Q(x_\shortparallel)\big|^2  
\,-\, \Omega_m^2  Q^2(x_\shortparallel) \right] \;.
\end{equation}

The interaction term ${\mathcal S}'$ may, on the other hand, be
conveniently written as follows:
\begin{equation}
	{\mathcal S}'(q, Q, \varphi ; {\mathbf r}(t)) \;=\; 
	 \int d^4x \, J(x) \, \varphi(x) 
	\;,\;\; J(x) \;\equiv \; J_e(x) \,+\, J_m(x)  
\end{equation}
where $J_e$ and $J_m$ are, respectively, concentrated on the atom and on
the medium. They are given by:
\begin{equation}
J_e(x) \;=\; g \, q(t) \,\delta^3({\mathbf x} - {\mathbf r}(t)) \;\;,\;\;\;
J_m(x) \;=\; \lambda \,Q(x_\shortparallel) \,\delta(x^3) \;,
\end{equation}
where $g$ and $\lambda$ are coupling constants.

\section{Transition amplitudes}\label{sec:tramp}
In order to study the transition amplitudes and probabilities which are
responsible for the quantum friction phenomenon, we adopt the interaction
picture, based on our choice for the free and interaction actions. In this
situation, we have the following expression for the time evolution of the 
operators corresponding to the dynamical variables:
\begin{align}\label{eq:fields}
\hat{q}(t) &=\; \frac{1}{\sqrt{2 m \Omega_e}} \, \big( \hat{a} e^{- i \Omega_e t}
\,+\, \hat{a}^\dagger e^{i \Omega_e t}  \big) \nonumber\\
\widehat{Q}(x_\shortparallel) &=\; \frac{1}{\sqrt{2 \Omega_m}} \, \big(
\hat{\alpha}({\mathbf x}_\shortparallel) \,  e^{- i \Omega_m t}
\,+\, \hat{\alpha}^\dagger({\mathbf x}_\shortparallel) \,  e^{i \Omega_m t} \big)
\nonumber\\
\widehat{\varphi}(x) &=\; 
\int \frac{d^3{\mathbf k}}{(2 \pi)^{3/2}} \frac{1}{\sqrt{2|{\mathbf k}|}} 
\big( \hat{a}({\mathbf k}) \,  e^{-i k \cdot x }
\,+\, \hat{a}^\dagger({\mathbf k}) \,  e^{i k \cdot x } \big) \;.
\end{align}
Here, the creation and annihilation operators satisfy the standard
commutation relations; namely, the only non-vanishing commutators are:
\begin{equation}
[\hat{a} \,,\, \hat{a}^\dagger] \,=\, 1 \;\;,\;\;
	[\hat{a}({\mathbf k}) \,,\, \hat{a}^\dagger({\mathbf p}) ] \,=\,
	\delta^3({\mathbf k} - {\mathbf p}) \;\;,
\end{equation}
and, taking into account the independence of the degrees of freedom for
different spatial points,
\begin{equation}
[\hat{\alpha}({\mathbf x}_\shortparallel) \,,\,
\hat{\alpha}^\dagger({\mathbf x}'_\shortparallel) ] \,=\, \delta^2({\mathbf
x}_\shortparallel - {\mathbf x}'_\shortparallel )
\;\;.
\end{equation}

In order to study quantum friction, we consider the usual situation of the
atom moving at a constant velocity, which is assumed to be parallel to
the plane. Without any loss of generality, we use coordinates such
that the velocity  points towards the $x^2$ direction. Analogously, also by
a proper choice of origin for space and time coordinates, the atom will
pass just above the origin of the plane at $t=0$. Denoting by $a$ the
(constant) distance between the atom and the plane, we then have:
\begin{equation}
	{\mathbf r}(t) \; = \; (0,\, v t , \, a ) \;.
\end{equation}

The transition amplitudes $T_{fi} = \langle f | \widehat{T} | i \rangle $
shall be determined from the scattering matrix $\widehat{S} = \hat{I} + i
\widehat{T}$, namely, from the evolution operator in the interaction
representation,	$\widehat{U}(t_f,t_i)$,  for $t_i \to - \infty$ and $t_f
\to  +\infty$: 
\begin{equation}\label{eq:U}
	\widehat{S} \;=\; \widehat{U}(+\infty,-\infty) \;=\; {\rm T}  \exp \big[ i
{\mathcal S}'(\hat{q}, \widehat{Q}, \widehat{\varphi}; {\mathbf r}(t))
\big]\;, 
\end{equation}
where ${\rm T}$ denotes time-ordering.

The initial quantum state $|i\rangle$ of the full system is assumed to be
the vacuum for all the modes, namely, for the electron, the medium, and the
vacuum field. In a self explanatory notation,
\begin{equation}\label{eq:in}
|i \rangle \;=\; |0_e \rangle \otimes |0_m \rangle \otimes |0_v \rangle \;.
\end{equation}
Regarding the final state, $|f\rangle$, in quantum friction there is no
production of vacuum-field particles (photons); indeed, that would require
a non-vanishing acceleration. Therefore, in quantum friction, only even
terms in the expansion of the exponential in (\ref{eq:U}) can intervene.
The lowest order contribution to the transition amplitude is, therefore,
the second order one, which yields:
\begin{equation}\label{eq:tfi}
	T_{fi} \;=\; i \int d^4x \int d^4x' \, 
		(\langle f_e | \otimes  \langle f_m  |) \, \widehat{J}_m(x)
		\widehat{J}_e(x') \,
		(|0_e \rangle \otimes |0_m \rangle ) \, G(x-x') \;,
\end{equation}
where we introduced the final states for the electron and the medium, and
the scalar field propagator $G$:
\begin{equation}
	G(x-x') \;=\; \int \frac{d^4k}{(2\pi)^4} \, e^{-i k \cdot (x-x')}
	\,\frac{i}{k^2 + i \varepsilon} \;,
\end{equation}
and we have assumed the initial and final states to be normalized. It is
rather straightforward to see that the only contribution to the transition
amplitude (to this order) contains a quantum for both $e$ and $m$, namely:
\begin{equation}\label{eq:states}
|f_e \rangle \;=\; \hat{a}^\dagger |0_e\rangle \;\;,\;\;\;
|f_m \rangle \;=\; \int d^2{\mathbf x}_\shortparallel \, f({\mathbf
x}_\shortparallel) \, \hat{\alpha}^\dagger({\mathbf x}_\shortparallel)
|0_m\rangle \;,
\end{equation}
where $\int d^2{\mathbf x}_\shortparallel \, | f({\mathbf
x}_\shortparallel) |^2 \,=\,1$.
Inserting this into (\ref{eq:tfi}), and integrating out $x$ and $x'$, we obtain:
\begin{equation}\label{eq:tfi1}
	T_{fi} \;=\; - \, \frac{2 \pi g \lambda}{\sqrt{2 m \Omega_e 2 \Omega_m}}
		\; \int \frac{d^3{\mathbf k}}{(2\pi)^3} \, 
		\delta(\Omega_e + \Omega_m + k_2 v) \,
	\frac{ e^{i k^3 a} \tilde{f^*}({\mathbf k}_\shortparallel)}{\Omega_m^2 - 
		{\mathbf k}^2 + i \varepsilon}
\end{equation}
where $\tilde{f}({\mathbf k}_\shortparallel) = \int d^2{\mathbf
x}_\shortparallel \, e^{-i {\mathbf k}_\shortparallel \cdot {\mathbf
x}_\shortparallel}f({\mathbf
x}_\shortparallel)$. Thus, integrating out $k^2$ and $k^3$, and denoting by
$k_x \equiv k^1$ the only remaining component to integrate, we have that
\begin{equation}\label{eq:tfi2}
T_{fi} \;=\; \frac{g \lambda}{4 v \sqrt{m \Omega_e \Omega_m}}
\; \int_{-\infty}^{+ \infty}   
\frac{d k_x}{2\pi} \, \tilde{f^*} \left( k_x, - \frac{\Omega_ e +
	\Omega_m}{v}\right) \, \frac{ e^{- a \,\sqrt{ k_x^2 +
	\Omega^2}}}{\sqrt{k_x^2 + \Omega^2}} \;,
\end{equation}
where we have defined 
\begin{equation}\label{eq:defOmega}
\Omega^2=\left(\frac{\Omega_e+\Omega_m}{v}\right)^2-\Omega_m^2 \;.
\end{equation}

\begin{figure}[!ht]
\centering
\includegraphics[scale=0.5]{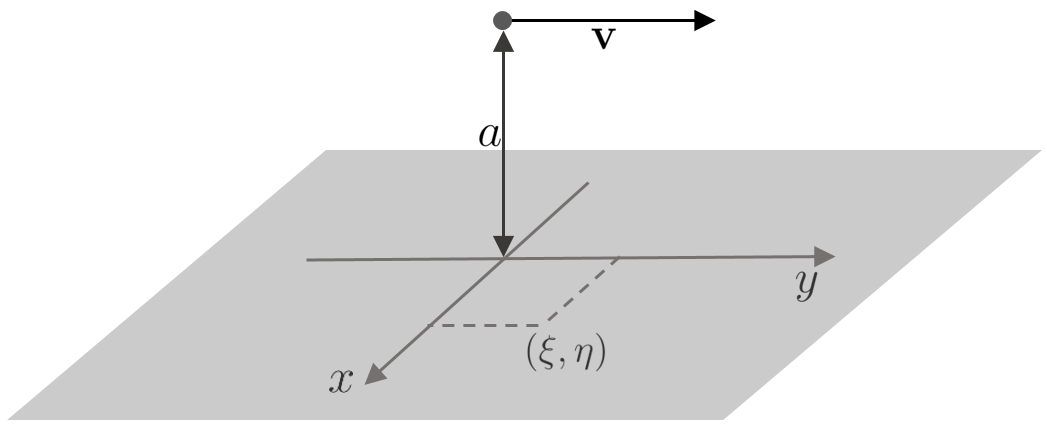}
\caption{Sketch of the system}
\label{fig:sketch}
\end{figure}

In what follows, for the sake of notational clarity, we use denote by $x$
and $y$ the two Cartesian coordinates $x^1$ and $x^2$, respectively.
We want to study the spatial properties of the transition probabilities, we
consider a spatially localized function $f$, centered about a point with
coordinates $(\xi, \eta)$ (see Fig.~\ref{fig:sketch}) on the plane, and with size $(\sigma_x,
\sigma_y)$:
\begin{equation}
	f(x,y) \,\equiv \, \phi_{\sigma_x}(x -
	\xi) \, \phi_{\sigma_y}(y - \eta)
\end{equation}
where
\begin{equation}
	\phi_{\sigma}(x) \,\equiv\, \frac{e^{- \frac{x^2}{4
	\sigma^2}}}{(2 \pi)^{1/4} \sqrt{\sigma}} \;\;\;,\;\;\;
	\int_{-\infty}^{+\infty} dx \, 	\abs{\phi_{\sigma}(x)}^2 \;=\; 1 \;.
\end{equation}
Then:
\begin{align}\label{eq:tfi3}
T_{fi} =&\; \frac{g \lambda}{v}\sqrt{\frac{\pi \sigma_x \sigma_y}{ 2 m \Omega_e
\Omega_m}}  \, e^{ - \sigma^2_y  \left(\frac{\Omega_ e +
\Omega_m}{v}\right)^2} \; e^{i\eta (\frac{\Omega_ e + \Omega_m}{v})} \nonumber\\
& \times \int_{-\infty}^{+ \infty} \,  
\frac{d k_x}{2\pi} \,  e^{- \sigma_x^2 k_x^2}
 \,  e^{- i k_x \xi}
	\frac{ e^{- a \,\sqrt{ k_x^2 +  \Omega^2}}}{\sqrt{k_x^2 +  \Omega^2}} \;.
\end{align}

Note that $\rho(\xi) \;\equiv\; |T_{fi}|^2/\sigma_x\sigma_y$ is a 
probability per unit area, and making $\sigma_x,\sigma_y\to0$ would give the probability
(per area) of having an oscillator of the medium excited at
$(\xi,\eta)$\footnote{This is equivalent to taking $f({\mathbf
x}_\shortparallel)\propto \delta(x-\xi)\delta(y-\eta)$ in (\ref{eq:states}).} 

\begin{equation}
 \rho(\xi)\equiv\lim_{\sigma_x,\sigma_y\to0}\frac{|T_{fi}|^2}{\sigma_x\sigma_y}=\frac{g^2\lambda^2}{8\pi m\Omega_m\Omega_e v^2}\abs{\int_{-\infty}^\infty dk_x~e^{-ik_x\xi }
	\frac{e^{- a \,\sqrt{ k_x^2 +  \Omega^2}}}{\sqrt{ k_x^2 +
	\Omega^2}}}^2 \;,
\end{equation}
which does not depend on $\eta$.

\begin{figure}[t]
    \centering
    \begin{tikzpicture}
        \begin{axis}[
            width=9cm,height=7.2cm,
            xmin=-0.1,
            xmax=0.1,
            ymin=0,
            ymax=+3.2,
            xlabel=$\Omega_e\xi$,
            ylabel=$\tilde{\rho}$,
            y label style={rotate=270},
            xtick={-0.1,-0.05,0.0,0.05,0.1},
            extra tick style={grid=major},
            legend style={at={(axis cs:0.02,1.2)},anchor=south west},
            legend cell align=left,
            y tick label style={
            /pgf/number format/.cd,
            fixed,
            precision=2,
            /tikz/.cd}
        ]
        \addplot+ [
            smooth,
            tension = 0.7,
            mark=none,
            ultra thick,
        ]
        file {001-1.dat};
        \addlegendentry{$\Omega_e a=0.01~,~f=1.0$},
        
        \addplot+ [
            smooth,
            blue,
            dashed,
            tension = 0.7,
            mark=none,
            ultra thick,
        ]
        file {002-1.dat};
        \addlegendentry{$\Omega_e a=0.02~,~f=1.0$},
        
        \addplot+ [
            smooth,
            orange,
            tension = 0.7,
            mark=none,
            ultra thick,
        ]
        file {001-15.dat};
        \addlegendentry{$\Omega_e a=0.01~,~f=1.5$},
        
        \addplot+ [
            smooth,
            orange,
            dashed,
            tension = 0.7,
            mark=none,
            ultra thick,
        ]
        file {002-15.dat};
        \addlegendentry{$\Omega_e a=0.02~,~f=1.5$},
        \end{axis}
    \end{tikzpicture}
    \caption{$\tilde{\rho}=\rho/\left(\frac{g^2\lambda^2}{4\pi
    m\Omega_m\Omega_e v^2}\right)$ as a function of $a$ (the distance
    between the atom and the medium), and the ratio between frequencies. 
    The four curves are for $v=0.1$, increasing the speed has
    a similar effect than increasing $f$.}
    \label{fig:distribution}
\end{figure}
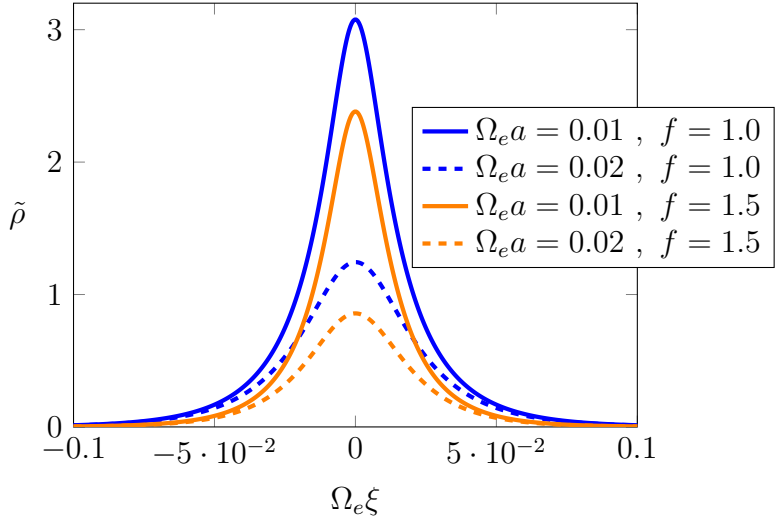

With  $\Omega$ as defined in (\ref{eq:defOmega}), and giving different
values to the adimensional combination $\Omega a$ for fixed $\Omega_e$, we
can see in the Fig.\ref{fig:distribution} how this distribution varies with
the distance between the medium and the atom, and with the quotient
$f\equiv \Omega_m/\Omega_e$.

\newpage

Integrating $\rho(\xi)$ for all $\xi$ and multiplying by a characteristic
length in the direction of movement of the atom, i.e. $vT$, where we can
think of $T$ as the time the atom has been moving with constant speed $v$,
gives the probability of this process to happen. Then, dividing by $T$,
would give the probability per unit time 
\begin{equation}\label{eq:P1}
    \mathcal{P}\equiv\frac{1}{T}~vT \int_{-\infty}^\infty d\xi~\rho(\xi)=\frac{g^2\lambda^2}{2mv\Omega_e\Omega_m}\int_0^\infty\,\, dk_x ~\frac{e^{-2a\sqrt{k_x^2 +  (\frac{\Omega_ e +\Omega_m}{v})^2 - \Omega_ m^2}}}{k_x^2 +  (\frac{\Omega_ e +\Omega_m}{v})^2 - \Omega_ m^2},
\end{equation}
which matches with the result obtained in equation (69) of \cite{Farias:2019lls}.

\section{Case of $u\neq 0$}

Now we consider the case where waves can be propagated through the medium at speed $u\neq0$, and see that results match with the previous model when $u\to0$. The action for the medium is now

\begin{equation}
    {\mathcal S}^{(0)}_m(Q) \,=\, 
	\int d^3x_\shortparallel \, \frac{1}{2} 
	\left[  \big(\partial_t Q(x_\shortparallel) \big)^2 \,-\, u^2 \,
	\big| {\mathbf \nabla}_\shortparallel Q(x_\shortparallel)\big|^2  
\,-\, \Omega_m^2  Q^2(x_\shortparallel) \right],
\end{equation}
and the second expression of (\ref{eq:fields}) turns into
\begin{equation}
    \hat{Q}(x_\shortparallel) =\; \int \frac{d^2{\mathbf k}_\shortparallel}{2 \pi} \frac{1}{\sqrt{2k_0}} \big( \hat{\alpha}({\mathbf k}_\shortparallel) \,  e^{-i k_\shortparallel \cdot x_\shortparallel }\,+\,\hat{\alpha}^\dagger({\mathbf k}_\shortparallel) \,  e^{i k_\shortparallel \cdot x_\shortparallel } \big), 
\end{equation}
where $k_0=\sqrt{u^2\abs{\mathbf{k}_\shortparallel}^2+\Omega_m^2}$ and $[\hat{\alpha}({\mathbf k}_\shortparallel) \,,\, \hat{\alpha}^\dagger({\mathbf p}_\shortparallel) ] \,=\,
\delta^2({\mathbf k}_\shortparallel - {\mathbf p}_\shortparallel)$. The (normalized) final state that we consider now is $\ket{f_m}=\frac{2\pi}{\ell}\alpha^\dagger({\mathbf p}_\shortparallel)\ket{0_m}$, i.e. an excitation with momentum $\mathbf{p}_\shortparallel$.\footnote{When $u=0$, this is equivalent to taking $f({\mathbf
x}_\shortparallel)\propto e^{-i\mathbf{p}_\shortparallel\cdot{\mathbf
x}_\shortparallel}$ in (\ref{eq:states}).} This gives a transition amplitude

\begin{align}\label{eq:tvel}
    T_{fi}=&\frac{2\pi}{\ell}\frac{g\lambda~e^{-a\sqrt{\abs{\mathbf{p}_\shortparallel}^2(1-u^2)-\Omega_m^2}}}{\sqrt{m\Omega_e\left(\abs{\mathbf{p}_\shortparallel}^2(1-u^2)-\Omega_m^2\right) \sqrt{u^2\abs{\mathbf{p}_\shortparallel}^2+\Omega_m^2}}}\nonumber\\
    &\times\delta\left(\Omega_e+\sqrt{u^2\abs{\mathbf{p}_\shortparallel}^2+\Omega_m^2}-v p_y\right).
\end{align}

The arising Dirac delta gives us some information about the process. First, $p_y$ has to be positive, so the momentum of the excitation in the medium has a positive component along the velocity of the atom. Then, it shows that there is a threshold for this process to occur: the speed of the atom should be greater than the speed of the waves propagating in the medium $v>u$.

\begin{equation}
    up_y<u\abs{\mathbf{p}_\shortparallel}<\sqrt{u^2\abs{\mathbf{p}_\shortparallel}^2+\Omega_m^2}=vp_y
\end{equation}

Dividing $\abs{T_{fi}}^2$ by $T$ with $u\to 0$ and integrating all possible momentums for the excitation of the medium gives the probability per unit time of having this process. 

\begin{equation}
    \mathcal{P}=\frac{1}{T}\left(\frac{\ell}{2\pi}\right)^2\int d^2\mathbf{p}_\shortparallel~\lim_{u\to0}\abs{T_{fi}}^2=\frac{g^2\lambda^2}{\pi mv\Omega_e\Omega_m}\int_0^\infty dp_x\frac{e^{-2a\sqrt{p_x^2 +  (\frac{\Omega_ e \Omega_m}{v})^2 - \Omega_ m^2}}}{p_x^2 +  (\frac{\Omega_ e +\Omega_m}{v})^2 - \Omega_ m^2},
\end{equation}
which is essentially the same  as (\ref{eq:P1}).

\section{Conclusions}\label{sec:conc}
In this paper, we have evaluated the transition amplitudes corresponding to
the elementary processes which  lead to the phenomenon of quantum friction
between a moving atom and a material plane. 
Our choice of system, and model, allows for a determination of the spatial
dependence of the Casimir friction phenomenon, by providing the funcional
form of the amplitude as a function of the distance, on the plane, to the
projection of the atom's trajectory.

The dependence of the previous results on the distance between the atom and
the plane is modulated by a precise combination of the model's parameters
and the velocity of the atom. 

Finally, by integrating out the probabilities to this order, we have found
agreement with the total vacuum decay probability obtained, for the
same model, by an evaluation of the imaginary part of the effective action,
in a functional integral approach.

\section*{Acknowledgements}
The authors thank ANPCyT, CONICET, CNEA and UNCuyo for financial support.

\end{document}